\RequirePackage{fix-cm}
\documentclass[twocolumn]{svjour3}          
\smartqed  
\usepackage{graphicx}
\usepackage{color}
 \usepackage{mathptmx}      

 \journalname{}
\begin{document}

\title{Survival Probability of the N\'eel State in Clean and Disordered Systems: an Overview}


\author{E. J. Torres-Herrera        \and
        Marco T\'avora \and
        Lea F. Santos 
}


\institute{E. J. Torres-Herrera \at
              Instituto de F{\'i}sica, Universidad Aut\'onoma de Puebla, Apt. Postal J-48, Puebla, Puebla, 72570, Mexico
           \and
            Marco T\'avora \at
              Department of Physics, Yeshiva University, New York, New York 10016, USA          
           \and
            Lea F. Santos \at
              Department of Physics, Yeshiva University, New York, New York 10016, USA \\ 
              ITAMP, Harvard-Smithsonian Center for Astrophysics, Cambridge, Massachusetts 02138, USA \\
              \email{lsantos2@yu.edu}                    
}

\date{}

\maketitle

\begin{abstract}
In this work we provide an overview of our recent results about the quench dynamics of one-\-dimensional many-body quantum systems described by spin-1/2 models. To illustrate those general results, here we employ a particular and experimentally accessible initial state, namely the N\'eel state. Both cases are considered: clean chains without any disorder and disordered systems with static random on-site magnetic fields. The quantity used for the analysis is the probability for finding the initial state later in time, the so-called survival probability. At short times, the survival probability may decay faster than exponentially, Gaussian behaviors and even the limit established by the energy-time uncertainty relation are displayed. The dynamics at long times slows down significantly and shows a powerlaw behavior. For both scenarios, we provide analytical expressions that agree very well with our numerical results.

\end{abstract}
\newpage
\section{Introduction}
\label{intro}
This overview describes our recent numerical and analytical results for the dynamics of quantum systems where external interactions with an environment are neglected~\cite{Torres2015,Torres2014PRA,Torres2014NJP,Torres2014PRAb,Torres2014PRE,TorresProceed,TorresKollmarARXIV}. The focus is on the effects of the internal interactions and on the interplay between interaction and disorder. The system is initially in a non-stationary state very far from equilibrium. Our main goal is to understand what characterizes the dynamics of such isolated many-body quantum system. 

Nonequilibrium quantum physics is a subject that permeates various fields of physics and chemistry, such as statistical mechanics, condensed matter physics, mole\-cular dynamics, quantum information, and cosmology. It is also intimately associated with the development of methods to control the dynamics of quantum systems,  aiming at slowing it down or accelerating it. The subject is, however, much less understood than equilibrium quantum physics. 

We have been continually emphasizing in our previous works that the dynamics of a quantum system~\cite{Torres2015,Torres2014PRA,Torres2014NJP,Torres2014PRAb,Torres2014PRE,TorresProceed,TorresKollmarARXIV} and also the new equilibrium reached by it~\cite{TorresKollmarARXIV,Torres2013} depend not only on its initial state or on its Hamiltonian but on both. The effects of the two are intertwined. The system evolution depends on where the energy of the initial state falls with respect to the spectrum of the Hamiltonian and on how much spread out this state is in the energy eigenbasis. Both aspects depend on the details of the Hamiltonian, such as its density of states and the presence or absence of disorder. 

To connect our analysis with experimental studies, it is convenient to think about the dynamics in terms of ``sudden quenches''. The system is initially in an eigenstate of an initial Hamiltonian $\widehat{H}_I$.  It is then taken far from equilibrium by a sudden perturbation (quench) that changes $\widehat{H}_I$ to a new final Hamiltonian $\widehat{H}_F$, initiating the evolution.
There is significant experimental interest in the quench dynamics of isolated many-body quantum systems. In nuclear magnetic resonance (NMR), for example, the initial state can be prepared by applying a magnetic field to the sample, which once turned off starts the dynamics~\cite{Batalhao2013}. In this context, a system of particular interest for us is the crystal of fluorapatite studied in solid state NMR~\cite{Cappellaro2007,Kaur2013}. The arrangement in this crystal is such that it can be treated for some time as a one-dimensional system of spins-1/2, which is similar to the systems considered in this work. Spin-1/2 models on a lattice can also be studied with trapped ions~\cite{Jurcevi2014,Richerme2014} and optical lattices~{\cite{Trotzky2008,Trotzky2012,Fukuhara2013}. The latter offer several advantages, including high controllability, quasi-isolation, and flexibility in the preparation of the initial state. 

A state that has received much attention in experiments with optical lattices is the N\'eel state, partially because of its importance in studies about magnetism. The state is such that the polarization of the spin on each site alternates along a chosen direction. We use this state for our illustrations below.

To quantify how fast the initial state $|\Psi(0)\rangle = |\mbox{ini}\rangle $ changes in time, we calculate the survival probability,
\begin{equation}
F(t) \equiv |\langle \mbox{ini} | e^{-i \widehat{H}_F t} |\mbox{ini} \rangle |^2  =  \left|\sum_{\alpha} |C_{\alpha}^{\small \mbox{ini}} |^2 e^{-i E_{\alpha} t}  \right|^2 ,
\nonumber \\
\label{eq:fidelity}
\end{equation}
where $C_{\alpha}^{\small \mbox{ini}} = \langle \mbox{ini} |\psi_{\alpha} \rangle$ is the overlap between the initial state and the eigenstates $|\psi_{\alpha} \rangle$ of the final Hamiltonian and $E_{\alpha}$ are the corresponding eigenvalues. The survival probability has received several different names, such as fidelity, non-decay probability, and return probability. It gives the probability for finding the initial state at time $t$. Notice that if we know the envelope of the energy distribution of the initial state weighted by the components $|C_{\alpha}^{\small \mbox{ini}} |^2$, that is, 
\begin{equation}
\rho_{\small \mbox{ini}}(E) = \sum_\alpha |C_{\alpha}^{\small \mbox{ini}}|^2\delta(E-E_\alpha) ,
\label{eq:LDOS}
\end{equation}
then by doing a Fourier transform we are able to obtain an analytical expression for $F(t)$. This distribution is often referred to as local density of states (LDOS) or strength function. We use the first designation, but the reader should not confuse LDOS with the density of states. The latter is simply the distribution of all eigenvalues of the Hamiltonian, without any reference to a specific initial state.

When dealing with unstable systems, such as unstable nuclei, the decay is exponential, as observed experimentally. However, deviations exist. They are associated with the following scenarios.

\subsection{Short Times} 
By expanding Eq.(\ref{eq:fidelity}), one sees that the initial decay has to be quadratic in time. This is the region of the quantum Zeno effect. 
\begin{equation}
F(t \rightarrow 0)  \approx 1-\sigma_{\small \mbox{ini}}^2 t^2,
\label{eq:quadratic}
\end{equation}
where
\begin{eqnarray}
\sigma_{\small \mbox{ini}} &&= \sqrt{\sum_{\alpha} |C_{\alpha}^{\small \mbox{ini}} |^2 (E_{\alpha} - E_{\small \mbox{ini}})^2} \nonumber \\
&&=\sqrt{\sum_{n \neq {\small \mbox{ini}} } |\langle n |\widehat{H}_F | \mbox{ini}\rangle |^2 },
\label{deltaE}
\end{eqnarray}
is the uncertainty in energy of the initial state and 
\begin{equation}
E_{\small \mbox{ini}} = \langle \mbox{ini} |\widehat{H}_F | \mbox{ini} \rangle = \sum_{\alpha} |C_{\alpha}^{\small \mbox{ini}}|^2 E_{\alpha} 
\label{Eini}
\end{equation}
is the energy of the initial state with respect to the final Hamiltonian. Note that $|n\rangle$ denotes the basis vectors used to write the Hamiltonian $\widehat{H}_F$, the initial state $| \mbox{ini} \rangle$ being one of them.

\subsection{Intermediate Times: Strong Perturbation} An exponential decay of the survival probability implies a Lorentzian LDOS. This is the regime of the Fermi golden rule. However, if the perturbation that takes $\widehat{H}_I$ into $\widehat{H}_F$ is sufficiently strong, being beyond perturbation theory, the decay of the survival probability can be faster than exponential~\cite{Torres2014PRA,Torres2014NJP,Torres2014PRAb,Torres2014PRE,TorresProceed,TorresKollmarARXIV}. 

In realistic systems with two-body interactions, as the ones treated here, the fastest decay for a unimodal LDOS is Gaussian. In this case, the decay rate coincides with the width $\sigma_{\small \mbox{ini}}$ of the LDOS: $F(t) = \exp(-\sigma_{\small \mbox{ini}}^2 t^2)$. Gaussian decays that continue beyond very short times were predicted for two-body random matrices (\cite{Izrailev2006} and references therein). The novelty of our studies is the verification that this Gaussian behavior can indeed emerge in realistic systems and that it can persist until $F(t)$ touches the saturation point. In addition, we have shown that for some classes of initial states, it is straightforward to find $\sigma_{\small \mbox{ini}}$ analytically~\cite{Torres2014PRA,Torres2014NJP}.

\subsection{Intermediate Times:  Bimodal LDOS} Another realistic case where the decay of the survival probability can be faster than exponential occurs when the LDOS is bimodal. In this scenario, the decay at short times can reach the fastest velocity determined by the energy-time uncertainty relation, while the behavior at later times depends on the shape of each peak~\cite{Torres2014PRAb}.

\subsection{Long Times and Disorder} 
Even if the system decays into the continuum, there is always a lower bound, $E_{cut}$, in the spectrum. Taking this bound into account, the Fourier transform of the LDOS necessarily leads to a decay slower than exponential at long times~\cite{Khalfin1958,Fonda1978}. There has been analytical~\cite{Campo2011,DelCampoARXIV,MugaBook} and experimental~\cite{Rothe2006} studies showing that the decay at long times should become powerlaw. The relation between the bounded energy spectrum and a powerlaw fidelity decay at long times is shown below for two simple cases. 

\subsubsection{Lorentzian LDOS} 
Let us write the survival probability as $F(t) = {\left| {A(t)} \right|^2}$. In the absence of a lower bound, when $\rho_{\rm{ini}} (E)$ is a Lorentzian of width $\Gamma_{\rm{ini}} $, the Fourier transform of the LDOS leads to the exponential decay of the survival probability, $\exp (- \Gamma_{\rm{ini}} t)$. We solve the integral,
\[
A(t) = \int_{-\infty}^{\infty} \frac{1}{2\pi} \frac{\Gamma_{\rm{ini}} }{(E_{\rm{ini}} - E)^2 +
 \Gamma_{\rm{ini}}^2 /4} e^{- i E t} dE ,
\]
 with residues. Since $t>0$, we close the contour clockwise on the lower plane, 
\begin{eqnarray}
&& \frac{\Gamma_{\rm{ini}} }{2\pi} \int_{-\infty}^{\infty} \frac{e^{- i E t}}{ [(E_{\rm{ini}} - E) + i \Gamma_{\rm{ini}} /2]   
 [(E_{\rm{ini}} - E) - i \Gamma_{\rm{ini}} /2] } dE \nonumber \\
 && = \frac{\Gamma_{\rm{ini}} }{2\pi} \oint \frac{ 
 \frac{e^{- i E t}}{   
 (E- E_{\rm{ini}} ) - i \Gamma_{\rm{ini}} /2}
  }{( E - E_{\rm{ini}} ) + i \Gamma_{\rm{ini}} /2]} \nonumber  .
\label{expPole1}
\end{eqnarray}

There is a pole at $E  = E_{\rm{ini}} - i \Gamma_{\rm{ini}} /2$. Taking into account the negative sign to $2\pi i$, we obtain
\begin{eqnarray}
&&   \frac{\Gamma_{\rm{ini}} }{2\pi}  (- 2\pi i)  \frac{e^{-i E_{\rm{ini}} t}  e^{- i (-i \Gamma_{\rm{ini}} /2) t}}{  - i  \Gamma_{\rm{ini}} } = e^{-i E_{\rm{ini}} t} e^{- \frac{\Gamma_{\rm{ini}} t}{2}}.
\label{expPole}
\end{eqnarray}

On the other hand, when the lower bound does exist, $A(t)$ can again be solved by replacing the integral with a contour integral in the complex plane~\cite{MugaBook}, but the contour is now as follows.
Choosing for convenience
the energy bound to be $E_{cut}=0$, the contour has the positive real energy axis, the arc of infinite radius running clockwise from the positive real energy axis to the negative imaginary axis, and the negative imaginary axis from $-i\infty$ to the origin, that is,
\begin{eqnarray}
 \oint\limits_{\cal C} \rho _{\rm{ini}}(E)\, e^{ - iEt} dE &=& \int_0^\infty  \rho _{\rm{ini}} (E) e^{ - iEt} dE \nonumber \\
 &+& \int_{arc}  \rho _{\rm{ini}} (E) e^{ - iEt} dE \nonumber \\
 &+& \int_{-i\infty}^0  \rho_{\rm{ini}} (E) e^{ - iE t} dE
\end{eqnarray}
In general, the contribution from the integration along the circular arc vanishes. Thus, using $E=-i\varepsilon$, we can write:
\begin{eqnarray}
A(t) &=& \int_0^\infty  {{\rho_{{\rm{ini}}}}} (E){e^{ - iEt}}dE\cr
&=&  \oint\limits_{\cal C} \rho_{\rm{ini}} (E)\, e^{ - iEt} dE + \int_0^{-i\infty}  \rho_{\rm{ini}} (E) e^{ - iE t} dE \cr
&=& \oint\limits_{\cal C} \rho_{\rm{ini}}(E)\,e^{ - iEt}dE - i\int_0^\infty  \rho_{\rm{ini}}( - i\varepsilon)\, e^{ - \varepsilon t} d\varepsilon \cr
&\equiv& {A_1}(t) + {A_2}(t) .
\end{eqnarray}
The first term, $A_1(t)$, depends on the poles in the fourth quadrant. As we saw in Eq.~(\ref{expPole}), it gives an exponential decay. It is the second term,
\begin{eqnarray}
{A_2}(t) &=&  - i\frac{{{\Gamma _{{\rm{ini}}}}}}{{2\pi }}\int_0^\infty  {d\varepsilon \,\frac{{{e^{ - \varepsilon t}}}}{{{{({E_{{\rm{ini}}}} + i\varepsilon)}^2} + \Gamma _{{\rm{ini}}}^2/4}}} , \nonumber
\end{eqnarray}
that leads to the powerlaw behavior at long times, because as $t$ becomes large, only small $\varepsilon$ survives, so we can set $\varepsilon =0$ at the denominator,
\begin{eqnarray}
{A_2}(t) &\mathop  \to \limits^{t \to \infty }&  - i\frac{{{\Gamma _{{\rm{ini}}}}}}{{2\pi }}\int_0^\infty  {d\varepsilon \,\frac{{{e^{ - \varepsilon t}}}}{{{{{E_{{\rm{ini}}}} }^2} + \Gamma _{{\rm{ini}}}^2/4}}} 
\propto \frac{1}{t} \nonumber
\end{eqnarray}
which implies that 
\begin{equation}
F(t \to \infty ) \propto \frac{1}{{{t^2}}}. 
\end{equation}

\subsubsection{Gaussian LDOS}
Another simple example is the Gaussian distribution, in which case
\begin{eqnarray}
A(t) &=&  \frac{1}{{\sqrt {2\pi \sigma _{{\rm{ini}}}^2} }}\int_0^\infty  {dE\,{e^{ - iEt}}{e^{ - {{( E - {E_{{\rm{ini}}}})}^2}/2\sigma _{{\rm{ini}}}^2}}} .
\end{eqnarray}

At very long times, the first exponential oscillates very fast, unless $E$ is very small. Similar to what we did in the Lorentzian case, we can then set $E=0$ for the second exponential,
\begin{eqnarray}
A(t)  
&\mathop  \to \limits^{t \to \infty }&  \frac{1}{{\sqrt {2\pi \sigma _{{\rm{ini}}}^2} }}\int_0^\infty  {dE\,{e^{ - iEt}}{e^{ - {{{E_{{\rm{ini}}}}}^2}/2\sigma _{{\rm{ini}}}^2}}} \propto \frac{1}{t}.
\end{eqnarray}
which again implies  $F(t \to \infty ) \propto \frac{1}{{{t^2}}}$. 

We note, however, that in our numerics for clean and disordered spin systems, the exponent of the powerlaw decay that we observe is $\le 1$. Therefore, for our systems, the main origin of the algebraic decay of $F(t)$ at long times is not the lower bound in the spectrum. Instead, this behavior is related to correlations between eigenstates. This is  further discussed in Section~4, in the context of a disordered spin chain~\cite{Torres2015}. 

This work is divided as follows. After describing the spin-1/2 models and the initial state considered  in Section~\ref{sec:MODEL}, we proceed to the two main parts of the paper. In Section~\ref{sec:FAST}, we present situations where the decay of the survival probability is faster than exponential. In Section~\ref{sec:POWERLAW}, we show a powerlaw decay that emerges at long times in disordered spin chains. Concluding remarks are provided in Section~\ref{sec:CONCLUSION}.


\section{Spin-1/2 Models and Initial States}
\label{sec:MODEL}

We consider a one-dimensional lattice of interacting spins-1/2 with an even number $L$ of sites and two-body interactions. The Hamiltonian is given by
\begin{eqnarray}
&&\widehat{H}= \sum_{k=1}^L h_k  \widehat{S}_k^z   +
\nonumber \\
&& + \sum_{k} J \left(
\widehat{S}_k^x \widehat{S}_{k+1}^x + \widehat{S}_k^y \widehat{S}_{k+1}^y +
\Delta \widehat{S}_k^z \widehat{S}_{k+1}^z \right) + \nonumber \\
&&+ \lambda \sum_{k} J \left(
\widehat{S}_k^x \widehat{S}_{k+2}^x + \widehat{S}_k^y \widehat{S}_{k+2}^y +
\Delta \widehat{S}_k^z \widehat{S}_{k+2}^z \right) \label{eq:Ham}  .
\end{eqnarray}
In the equation above, we consider the Planck constant $\hbar =1$, $\widehat{S}^{x,y,z}_k $ are spin operators, $h_k$ are the Zeeman splittings of each site $k$, $J=1$ sets the energy scale, and $\Delta$ is the anisotropy parameter. The sums in the second and third lines run from $k=1$ to $L$ for periodic boundary conditions and from $k=1$ to $L-1$ for open boundary conditions. The total spin in the $z$-direction, $\widehat{{\cal{S}}}^z=\sum_k\widehat{S}_k^z$, is conserved. We work with the largest subspace, ${\cal{S}}^z=0$, of dimension ${\cal N}=L!/(L/2)!^2$. 

For the clean system investigated here, the magnetic field along the $z$ axis is constant, $h_k= h$. The disordered system is characterized by random static magnetic fields, where the amplitudes $h_k$ are random numbers from a uniform distribution $[-h,h]$. We also look at the case where a single defect exists, that is, only one site has $h_k$ different from the others.

In the absence of disorder ($h_k= h$) and of next-nearest-neigh\-bor (NNN) couplings ($\lambda=0$), the Hamiltonian is solvable with the Bethe ansatz~\cite{Bethe1931}. This clean Hamiltonian with only nearest-neighbor (NN) couplings is referred to as XXZ model. Disorder~\cite{Avishai2002,SantosEscobar2004,Dukesz2009}, even a single defect~\cite{Santos2004,Gubin2012}, or NNN couplings~\cite{Gubin2012,Santos2008PRE,Santos2009JMP} can take the system into the chaotic domain.

\paragraph{Initial State.}
We take as initial state one of the so-called site-basis vectors (also known as computational basis vectors). They correspond to states that are on-site localized, so the spin on each site either points up in the $z$-direction or down. The initial state selected is the N\'eel state, which is given by
\begin{equation}
|\mbox{NS} \rangle = |\uparrow \downarrow \uparrow \downarrow \uparrow \downarrow \uparrow \downarrow \uparrow \downarrow  \ldots \rangle .
\label{eq:NEEL}
\end{equation}

The scenario where a site-basis vector evolves according to the Hamiltonian given by Eq.~(\ref{eq:Ham}) is equivalent to a quench where the initial Hamiltonian is the Ising part of the total Hamiltonian, $\widehat{H}_I = \sum_{k} \widehat{S}_k^z \widehat{S}_{k+1}^z $, and the final Hamiltonian is  $\widehat{H}_F =\widehat{H}$ from Eq.~(\ref{eq:Ham}). The perturbation to change the Ising $\widehat{H}_I $ to $\widehat{H}_F$, where couplings in the $xy$-plane also exist, is very strong, being beyond perturbation theory. As a consequence, the initial decay of the survival probability is expected to be faster than exponential.

When the initial state is a site-basis vector, it is straightforward to obtain its energy $E_{\small \mbox{ini}}$ [Eq.~(\ref{Eini})] and its energy uncertainty $\sigma_{\small \mbox{ini}}$  [Eq.~(\ref{deltaE})]. 
As indicated by those equations, the energy is simply the diagonal element of the Hamiltonian matrix written in the site-basis and  $\sigma_{\small \mbox{ini}}^2$ is the sum of the squares of the off-diagonal elements in the row of that chosen initial state. To obtain $\sigma_{\small \mbox{ini}}^2$, we just need to count how many site-basis vectors are directly coupled to the chosen initial state.


\section{Clean Systems: Faster than Exponential Decays}
\label{sec:FAST}

We analyze two examples where the decay of the survival probability is faster than exponential: when the LDOS is a single Gaussian and when it is composed of two well separated Gaussians.

\subsection{Gaussian Decay}
Let us analyze first the case where the final Hamiltonian describes an anisotropic clean system with open boundary conditions, NN and NNN couplings: $h_k=h$, $\Delta=1/2$, $\lambda=1$. The density of states, as shown in Fig.~\ref{fig:FAST}  (a), has a Gaussian shape. This is typical of systems with two-body interactions \cite{Brody1981,Kota2001}, although the distribution is not necessarily symmetric~\cite{Santos2012PRE,Zangara2013}. The Gaussian form implies that most states concentrate in the middle of the spectrum. This is the region where strong mixing can occur and where the eigenstates are therefore more delocalized. As a consequence, an initial state with energy $E_{\small \mbox{ini}} $ (\ref{Eini}) close to the center of the spectrum decays faster than a state with $E_{\small \mbox{ini}} $ closer to the edges.

\begin{figure*}
\center
\includegraphics*[width=0.7\textwidth]{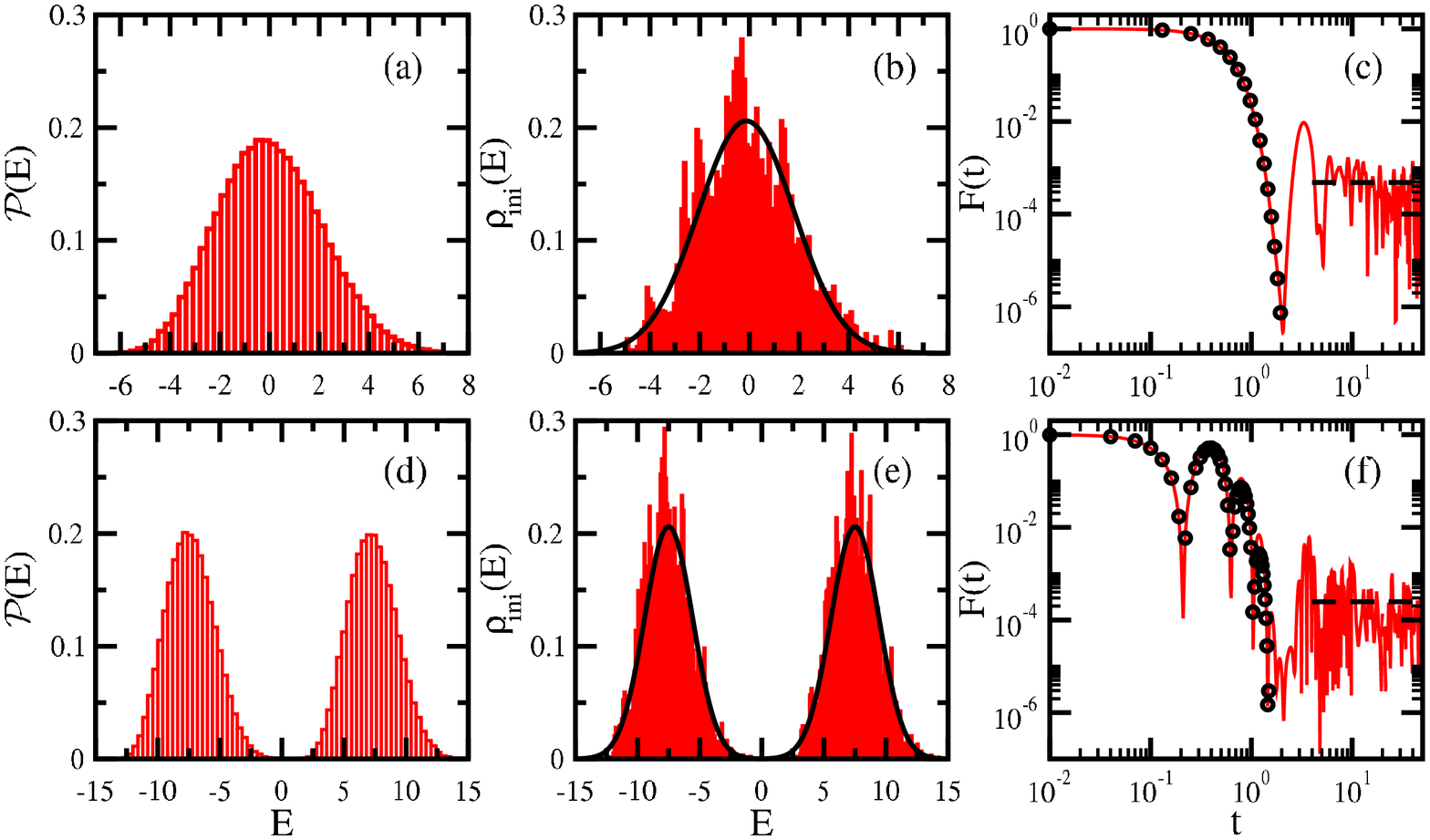}
\caption{Density of states (a) and (d); LDOS (b) and (e); and survival probability decay (c) and (f). Open boundary conditions; $L=16$. Clean Hamiltonian with NN and NNN couplings, $\Delta=1/2$, $\lambda=1$ (a,b,c) and the same Hamiltonian with an additional defect $\epsilon=15$ on site $L/2$ (d,e,f). Initial state is the N\'eel state on (b) and (c) and an equally weighted superposition of two N\'eel states on (e) and (f). The solid lines in (b) and (e) are the Gaussian distributions with width (\ref{eq:WneelClean}).
In (c) and (f), the solid lines are the numerical results, while the circles correspond to the
analytical expressions,  Eq.~(\ref{eq:GaussF}) and Eq.~(\ref{eq:TG}), respectively. Dashed lines give the saturation points [Eq.(\ref{Eq:saturationF})].}
\label{fig:FAST}       
\end{figure*}

The initial state considered is the N\'eel state [Eq.~(\ref{eq:NEEL})].
Its LDOS is also Gaussian as seen in Fig.~\ref{fig:FAST}  (b).  This is an indication that this initial state has reached its maximum possible spreading, which is limited by the density of states. Such maximum spreading is due to the strong perturbation that takes $\widehat{H}_I $ to $\widehat{H}_F$. The  LDOS is centered at the energy obtained from  Eq.~(\ref{Eini}), which for the N\'eel state is
\begin{equation}
E_{\small \mbox{NS}}^{\small \mbox{open,clean}} =  \frac{ J\Delta}{4}  [ -(L-1) + (L-2)\lambda ] .
\label{eq:EneelClean}
\end{equation}
The width of the LDOS corresponds to the uncertainty in the energy of the initial state [Eq.~(\ref{deltaE})], which for our initial state is
\begin{equation}
\sigma_{\small \mbox{NS}}^{\small \mbox{open,clean}}  = \frac{ J}{ 2} \sqrt{L-1} .
\label{eq:WneelClean}
\end{equation}
For the chosen parameters of the final Hamiltonian, $E_{\small \mbox{NS}}^{\small \mbox{open,clean}} $ is close to the middle of the spectrum, which explains why the LDOS is well filled. For states with energy closer to the edges of the spectrum, the LDOS becomes more sparse. Some examples of this latter case may be found in \cite{Torres2014PRA,Torres2014NJP,Torres2014PRAb,Zangara2013}.

The decay of the survival probability is shown in Fig.~\ref{fig:FAST}  (c). The behavior is Gaussian (solid line) all the way to the saturation point (dashed line) and it agrees very well with the analytical expression (circles), 
\begin{equation}
F(t) = \exp (-\sigma_{\small \mbox{ini}}^2 t^2), 
\label{eq:GaussF}
\end{equation}
with $\sigma_{\small \mbox{ini}} $ given by Eq.~(\ref{eq:WneelClean}).

The decay of the survival probability eventually saturates because we deal with finite systems. In a system without too many degeneracies, the off-diagonal terms in the second term on the left-hand side of the equation
\[
F(t) = \sum_{\alpha} |C_{\alpha}^{\mbox{\small ini}} |^4 + \sum_{\alpha \neq \beta} |C_{\alpha}^{\mbox{\small ini}} |^2 |C_{\beta}^{\mbox{\small ini}} |^2 e^{i (E_{\alpha} - E_{\beta}) t} 
\]
averages out, leading to the infinite time average, 
\begin{equation}
\overline{F}=\sum_{\alpha} |C_{\alpha}^{\mbox{\small ini}} |^4 \equiv \mbox{IPR}_{\mbox{\small ini}} .
\label{Eq:saturationF}
\end{equation}

One sees that the saturation point depends only on how much spread out the initial state is in the energy eigenbasis. If many eigenstates $|\psi_{\alpha}\rangle$ contribute to the evolution of $|\mbox{ini}\rangle$, then there are many very small components $|C_{\alpha}^{\mbox{\small ini}} |^4$ and the saturation point is low.

In Eq.~(\ref{Eq:saturationF}), IPR stands for inverse participation ratio. This is one of the most common quantities used to measure the level of delocalization of an arbitrary state $|\xi \rangle = \sum_j C_j |\phi_j \rangle$ in a certain basis $|\phi\rangle$, 
\begin{equation}
\mbox{IPR}_{|\xi \rangle} = \sum_j |C_j|^4 .
\end{equation}
When the state coincides with one of the basis vectors, then IPR=1. The more delocalized the state is in the chosen basis, the smaller the value of IPR is. The minimum value for systems with time reversal invariance is reached by the eigenstates of real and symmetric full random matrices. In this case, the states are random vectors, leading to IPR$\sim3/{\cal N}$ \cite{ZelevinskyRep1996}.

\subsection{Cosine Square Decay: Lower Bound from the Energy-Time Uncertainty Relation}
We consider again a final Hamiltonian with NN and NNN couplings, open boundary conditions, $\Delta=1/2$, $\lambda=1$, but now not all sites have  $h_k=h$. Site $L/2$ is the defect site with $h_{L/2} = h + \epsilon$, where $\epsilon$ is an excess on-site energy. For $\epsilon \gg 1$, the defect effectively breaks the chain in two. The spectrum ends up having two sets of eigenvalues. One set corresponds to the states that do not have an excitation on the defect site, so they have low energies, and the other consists of the states with an excitation on the defect site, so they have large energies. The resulting density of states is a bimodal distribution with two well separated Gaussians, as shown in Fig.~\ref{fig:FAST}  (d).

To study the effects of these two peaks on the dynamics, we choose as initial state a superposition of two N\'eel states, so that one has an excitation on the defect and the other does not,
\begin{equation}
|\mbox{NS} \rangle_{sup} = \frac{|\uparrow \downarrow \uparrow \downarrow \uparrow \downarrow \uparrow \downarrow \uparrow \downarrow  \ldots \rangle  +  | \downarrow  \uparrow \downarrow \uparrow \downarrow \uparrow \downarrow \uparrow \downarrow \uparrow \ldots \rangle}{\sqrt{2}} .
\label{eq:superNEEL}
\end{equation}

The LDOS for this superposition is shown in Fig.~\ref{fig:FAST}  (e). As expected, it is also bimodal with one peak centered at $E_{\small \mbox{NS}}^{\small \mbox{open,clean}} -\epsilon/2$ and the other at $E_{\small \mbox{NS}}^{\small \mbox{open,clean}}+\epsilon/2$. The peaks have the same widths given by $\sigma_{\small \mbox{NS}}^{\small \mbox{open,clean}}$ [(Eq.~\ref{eq:WneelClean})]. 

The decay of the survival probability, shown in  Fig.~\ref{fig:FAST}  (f), agrees very well with the Fourier transform of the two Gaussian peaks,
\begin{equation}
F(t) = \cos^2\left(\frac{\epsilon}{2} t\right)\exp\left[-(\sigma_{\small \mbox{NS}}^{\small \mbox{open,clean}})^2 t^2\right].
\label{eq:TG}
\end{equation}

The decay at short time is dominated by the cosine square part of Eq.~(\ref{eq:TG}), which later leads to revivals. The envelope of these subsequent damped oscillations is determined by the Gaussian part of Eq.~(\ref{eq:TG}).

\paragraph{Lower Bound Established by the Energy-Time Uncertainty Relation.}
The cosine square decay during $t <\pi/\epsilon$  is the fastest decay of the survival probability allowed by the energy-time uncertainty relation.

The lower bound for the decay\cite{Mandelstam1945,Ersak1969,Fleming1973,Bhattacharyya1983,Gislason1985,Vaidman1992,Uffink1993,Pfeifer1993,GiovannettiPRA2003,Boykin2007},
\begin{equation}
F(t) \geq \cos^2(\sigma_{\mbox{\small ini}} t) ,
\label{CosLimit}
\end{equation}
can be derived from the Mandelstam-Tamm uncertainty relation,
\[
\sigma_H \sigma_A \geq \frac{1}{2} \left| \frac{d\langle \widehat{A} \rangle}{dt} \right| .
\]
The uncertainty in energy, $\sigma_H $, for a non-stationary state coincides with $\sigma_{\mbox{\small ini}}$.
If $\widehat{A}$ is the projection operator on the initial state, $\widehat{A}=|\mbox{ini}\rangle \langle \mbox{ini}|$, then
$\langle \widehat{A} \rangle = F(t)$ and 
$\sigma_A^2 = F(t) - F(t)^2$.
Thus
\[
\sigma_{\mbox{\small ini}} \sqrt{F(1-F)} \geq \frac{1}{2} \left| \frac{dF}{dt} \right|.
\]
To solve this equation we can follow Ref.~\cite{Uffink1993} and  write $F(t)=\cos^2\theta$, so that 
\begin{eqnarray}
&&\sigma_{\mbox{\small ini}} \cos\theta \sin\theta   \geq \frac{1}{2} \left| -2 \cos\theta \sin\theta \frac{d\theta}{dt} \right| \nonumber \\
&&
\Rightarrow   \left|\frac{d\theta}{dt} \right| \leq \sigma_{\mbox{\small ini}} 
\Rightarrow  \left| \theta(t) \right| \leq \sigma_{\mbox{\small ini}} t \nonumber \\
&&
\Rightarrow \left| \arccos \left( \sqrt{F(t)} \right) \right|   \leq \sigma_{\mbox{\small ini}} t . \nonumber
\end{eqnarray}
Since the $\arccos$ is a strictly decreasing function,
\[
 \arccos \left( \sqrt{F(t)} \right)  \geq \sigma_{\mbox{\small ini}} t \Rightarrow \sqrt{F(t)} \geq \cos(\sigma_{\mbox{\small ini}} t),
 \]
from where Eq.~(\ref{CosLimit}) follows. Notice that this expression is valid only while the cosine is positive. In our case, where the LDOS is composed of two well separated peaks, $\sigma_{\mbox{\small ini}} \sim \epsilon/2$.

\section{Disordered Systems: Powerlaw Decays}
\label{sec:POWERLAW}
For the relatively small clean systems studied above, it is not straightforward to identify the powerlaw decay at long times. In the case of Fig.~\ref{fig:FAST}  (c), for instance, we need to do a time average of $F(t)$ to notice that there is indeed an algebraic decay for $2 < t < 100$, before the the survival probably starts to simply fluctuate around its infinite time average. The powerlaw exponent of this decay is very close to 1, that is $F(t) \propto t^{-1}$.

To explore the onset of smaller powerlaw exponents, we now add disorder to the system, which slows down its dynamics. We consider the disordered XXZ model described by an anisotropic Hamiltonian ($\Delta=0.48$) with periodic boundary conditions, NN couplings only ($\lambda=0$), and random static magnetic fields, where $h_k$ are random numbers uniformly distributed in $[-h,h]$. 

In a non-interacting system ($\Delta=0$), the presence of disorder localizes the excitations. This is the scenario of the Anderson localization~\cite{Anderson1958}, which has been extensively studied~\cite{Kramer1993,Evers2008,Izrailev2012} and also tested experimentally, more recently in two-dimensional ultracold gases with speckle disorder~\cite{MorongARXIV}. 

It had been conjectured that localization should persist also when interactions would be taken into account \cite{Anderson1958,Fleishman1980}. This was confirmed with perturbative arguments in~\cite{Gornyi2005,Basko2006} and rigorously in~\cite{ImbrieARXIV}. So far, this so-called many-body localization (MBL) has been tested in one experiment with cold atoms in optical lattices~\cite{SchreiberARXIV}. Here, we study the evolution of the N\'eel state as the system approaches the MBL phase.

When $h\sim1$, the disordered XXZ model is chaotic \cite{Avishai2002,SantosEscobar2004,Dukesz2009,Santos2004}, so the eigenstates close to the middle of the spectrum are very delocalized in the site-basis, being similar to random vectors. Their inverse participation ratio is inversely proportional to the dimension of the symmetry sector to which they belong. Analogously, initial states corresponding to site-basis vectors with energy close to the center of the spectrum of a chaotic $\widehat{H}_F$ have $\mbox{IPR}_{\mbox{\small ini}} \propto {\cal N}^{-1}$. 

As $h$ increases, the eigenstates $|\psi_{\alpha}\rangle$ become more localized, sampling only a portion of the Hilbert space, so $\mbox{IPR}_{\alpha} \propto {\cal N}^{-D_2}$, where $D_2<1$. Equivalently, the initial site-basis states get less spread out and $\mbox{IPR}_{\mbox{\small ini}} \propto {\cal N}^{-\tilde{D}_2}$ with $\tilde{D}_2<1$. The exponents $D_2$ and $\tilde{D}_2$ are known as generalized dimensions. For one-dimensional systems without interaction, they coincide,  $D_2= \tilde{D}_2$,  as shown in Ref.~\cite{Huckestein1997}. We verified that the same holds for our interacting system~\cite{Torres2015}.

\paragraph{Anderson Localization and Powerlaw Decay.} In studies of the Anderson localization, it has been shown that the decay of the survival probability averaged over random realizations, $\langle F(t)\rangle$, becomes powerlaw at the critical point, with an exponent that coincides with $\tilde{D}_2$ \cite{Ketzmerick1992,Huckestein1994,Huckestein1999}. This generalized dimension is obtained from the scaling analysis of the inverse participation ratio $\mbox{IPR}_{\mbox{\small ini}}$.

The subscript ``2'', appearing in $D_2$ and $\tilde{D}_2$, is used to distinguish from generalized dimensions obtained from scaling analysis of $\sum_{\alpha} |C_{\alpha}^{\mbox{\small ini}}|^{2 q}$ with $q\neq 2$, which are not considered in this work. They are important when investigating multifractal features of the states.

The coincidence between the powerlaw exponent and $\tilde{D}_2$ may be understood from the expression of the survival probability as follows. Introducing the identity
\begin{eqnarray}
{e^{i({E_\alpha } - {E_\beta })t}} = \int_{ - \infty }^\infty  {d\omega \,{e^{i\omega t}}\delta ({E_\alpha } - {E_\beta } - \omega )}
\end{eqnarray}
into $\left\langle {F(t)} \right\rangle$  from Eq.~(\ref{eq:fidelity}), we obtain:
\begin{eqnarray}
\langle F(t)\rangle  &=& \left\langle {\sum\nolimits_{\alpha ,\beta } {{{| {C_{\rm ini}^\beta } |}^2}{{| {C_{\rm ini}^\alpha }|}^2}{e^{i({E_\beta } - {E_\alpha })t}}} } \right\rangle \cr
 &\equiv& \int_{ - \infty }^\infty  {d\omega \,{e^{i\omega t}}} C(\omega ) ,
 \label{eq24}
\end{eqnarray}
where $C(\omega )$ is the correlation function given by~\cite{Kratsov2011},
\begin{eqnarray}
C(\omega ) \equiv \left\langle {\sum\nolimits_{\alpha ,\beta } {{{| {C_{{\rm{ini}}}^\beta } |}^2}{{| {C_{{\rm{ini}}}^\alpha } |}^2}\delta ({E_\alpha } - {E_\beta } - \omega )} } \right\rangle .
\end{eqnarray}
The correlation function $C(\omega )$ quantifies the overlap between two eigenstates with energy difference $\left| {{E_\alpha } - {E_\beta }} \right|$. In the critical regime, $C(\omega)$ decays as a powerlaw~\cite{Huckestein1994,ChalkerDaniell1988}; for small $\omega$ and large $\mathcal{N}$, it scales as~\cite{Huckestein1994,Cuevas2007},
\begin{eqnarray}
C(\omega  \to 0) \propto \frac{1}{\cal N}{\omega ^{{{\tilde D}_2} - 1}} .
\label{Comega}
\end{eqnarray}
At large $t$, due to the rapidly oscillating exponential term in Eq.~(\ref{eq24}), the integral
is dominated by small $\omega$. If we then
substitute Eq.~(\ref{Comega}) into Eq.~(\ref{eq24}), we obtain
\begin{eqnarray}
\langle F(t)\rangle  \propto {t^{ - {{\tilde D}_2}}} .
\end{eqnarray}
We see that $\tilde{D}_2$ measures how much correlated the components of the initial state, $|C^{\alpha}_{\mbox{\small ini}} |^2$, are and equivalently how much correlated the eigenstates are. In a chaotic system, where the states are uncorrelated random vectors,  $\tilde{D}_2 \sim 1$ and $C(\omega  \to 0) \propto \frac{1}{\cal N}$. As the disorder increases, the states shrink, correlations build up, and $\tilde{D}_2$ becomes smaller than 1.

We note, however, that the comparisons between the generalized dimension and the powerlaw exponent have in fact been carried out for the time-averaged survival probability, defined as 
\begin{equation}
C(t) \equiv \frac{1}{t} \int_0^t \langle F(\tau) \rangle d\tau  .
\end{equation}
This is a way to smooth the curve of the survival probability.  In order to reduce also the fluctuations in the values of IPR$_{\mbox{\small ini}}$, the scaling analysis is often done with the so-called typical inverse participation ratio, IPR$^{\mbox{\tiny typ}} \equiv \exp (\langle \ln\mbox{IPR}_{\mbox{\small ini}}\rangle)$. The scaling analysis of IPR$^{\mbox{\tiny typ}}$ gives $\tilde{D}_2^{\mbox{\tiny typ}}$.

\paragraph{Many-body localization and powerlaw decay.} Recently, we verified that the correspondence between the powerlaw exponent and the generalized dimension holds also in the presence of interactions~\cite{Torres2015}. The agreement is excellent for both $C(t)$ and $\langle F(t)\rangle$ when the disorder is small and the system sizes are large. As the disorder increases, the largest system sizes available for exact diagonalization, $L=14, 16$, are still too small. In this case,  oscillations are observed before the powerlaw behavior, with $\tilde{D}_2$ now capturing the decay of these oscillations. There is agreement between $\tilde{D}_2^{\mbox{\tiny typ}}$ and the powerlaw exponent of $C(t)$, but for a  short time interval.

In Ref.~\cite{Torres2015}, we performed averages over realizations and also over initial states. The latter was a set of site-basis vectors with the closest energies to the middle of the spectrum. In the present work, the initial state is always the N\'eel state, so the averages are performed only over realizations.
The averaged energy of the N\'eel state is
\begin{eqnarray}
\langle E_{\mbox{\small NS}}^{\mbox{\small closed, diso}}\rangle &=&-\frac{ J\Delta}{4}L +\left\langle\sum_{k=1}^{L}(-1)^{k}h_k\right\rangle \nonumber \\
&\sim& -\frac{ J\Delta}{4}L .
\label{eq:NSclosed}
\end{eqnarray}

\begin{figure}[!b]
\center
\includegraphics*[width=0.5\textwidth]{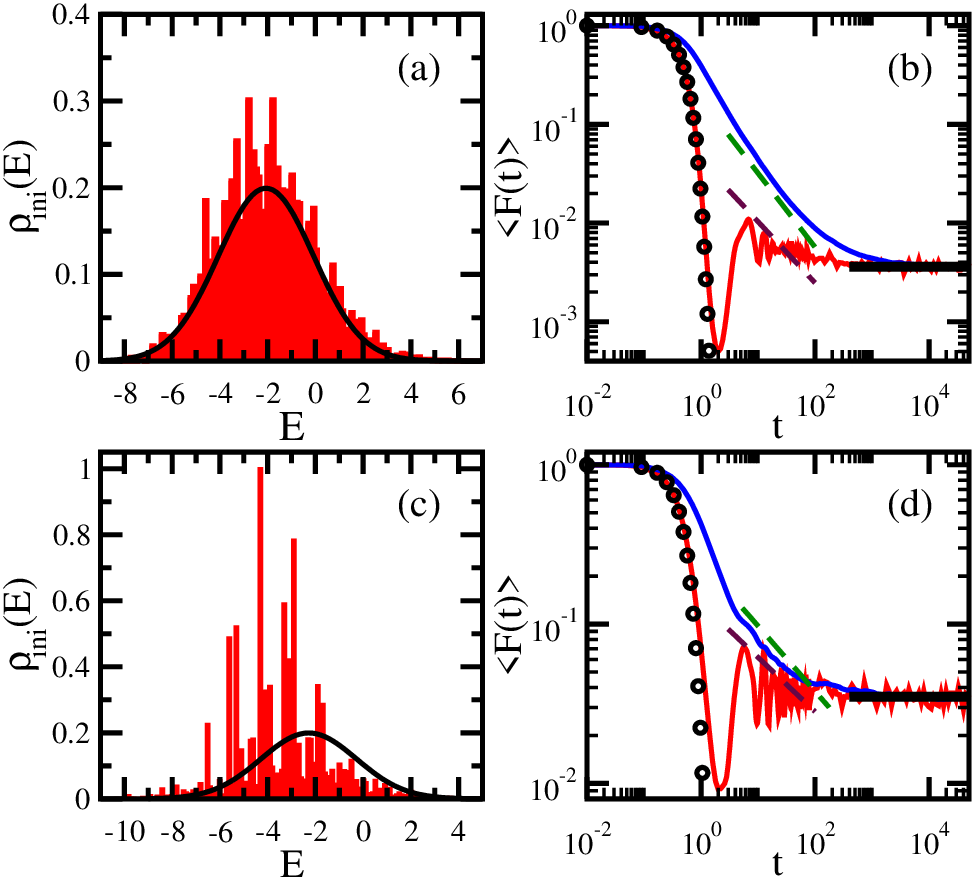}
\caption{LDOS for one disorder realization (a) and (c) and survival probability averaged over $10^3$ disorder realizations (b) and (d) for $h=1.5$ (a,b) and $h=3$ (c,d). Periodic boundary conditions; $\Delta=0.48$, $\lambda=0$, $L=16$. Initial state is the N\'eel state. Solid lines in (a) and (c) are Gaussians centered at $\langle E_{\mbox{\tiny NS}}^{\mbox{\tiny closed, diso}}\rangle$ [Eq.~(\ref{eq:NSclosed})] with width given by Eq.(\ref{eq:Wclosed}). In (b) and (d), the bottom solid curve gives $\langle F(t)\rangle$ and the top solid curve corresponds to $C(t)$.  The bottom dashed line gives $t^{-\tilde{D}_2}$ and top dashed line gives $t^{-\tilde{D}_2^{\mbox{\tiny typ}}}$. Circles indicate the Gaussian decay: $\exp\left[- (\sigma_{\mbox{\tiny NS}}^{\mbox{\tiny closed, diso}} )^2 t^2 \right]$. Thick horizontal lines represents the saturation points.}
\label{fig:POWER}       
\end{figure}

This value is still far from the edge of the spectrum,  but it is not at the center of the spectrum either, so it should be easier to localize this state than the states that we considered in~\cite{Torres2015}. Notice that it is the absence of NNN couplings that pushes the energy of the N\'eel state away from the middle of the spectrum [compare the equation above with Eq.~(\ref{eq:EneelClean})].

In Figs.~\ref{fig:POWER} (a), (c), we show the LDOS for one disorder realization for two values of the disorder strength, $h=1.5$ and $h=3$, respectively. The distribution becomes visibly more sparse as $h$ increases, although the width does not change, since it is independent of $h$,
\begin{equation}
\langle\sigma_{\mbox{\small NS}}^{\mbox{\small closed, diso}}\rangle=\frac{J}{2} \sqrt{L}.
\label{eq:Wclosed}
\end{equation}
As a consequence, the short-time dynamics of the survival probability, which depends only on the width of the LDOS, is identical for both values of $h$, as shown in Figs.~\ref{fig:POWER} (b) and (d). 

The bottom solid curve in Figs.~\ref{fig:POWER} (b) and (d) indicates $\langle F(t)\rangle$ and the top solid one represents the average $C(t)$. Both $\langle F(t)\rangle$,  and $C(t)$ reach the same saturation point and this values naturally increases as $h$ increases.

The initial Gaussian decay of $\langle F(t)\rangle$  is followed by damped oscillations and then finally saturation. It is interesting that at short times, around $t\sim2$, $\langle F(t)\rangle$ reaches values significantly lower than the saturation point. This is a result that deserves further investigation. Site-basis states that, like the N\'eel state, overshoot their decay seem to be behind the emergence of time intervals where $\langle F(t)\rangle$ remain below the saturation point, as it  observed in Fig.~1 of Ref.~\cite{Torres2015}.

In Figs.~\ref{fig:POWER} (b) and (d), the algebraic decay $t^{-\tilde{D}_2}$ (the bottom dashed line that is closest to $\langle F(t)\rangle$) matches the envelope of the decay of the oscillations of $\langle F(t)\rangle$. This occurs even for $h=1.5$, which is in contrast with our results in~\cite{Torres2015}, where the oscillations were minor for small disorder and $\tilde{D}_2$ agreed very well with the powerlaw exponent. The reason for this discrepancy is that even close to the chaotic regime of the disordered XXZ Hamiltonian, the N\'eel state is not as delocalized as the states we studied in \cite{Torres2015}, because of its low energy, so oscillations appear already at small $h$. In fact, for this disordered system, the N\'eel state does not seem to be able to reach a diffusive behavior, where $\langle F(t)\rangle \propto t^{-1}$, having instead $\tilde{D}_2, \tilde{D}_2^{\mbox{\tiny typ}} <1$ for all values of $h>1$ [see also Fig.~\ref{fig:D2} (d)].

The oscillations of $\langle F(t)\rangle$ are substituted by a powerlaw behavior when the average  $C(t)$ is considered [Figs.~\ref{fig:POWER} (b) and (d)]. Despite holding for a relatively short time interval, there is reasonable agreement between the numerical curve and $t^{-\tilde{D}_2^{\mbox{\tiny typ}}}$ (the latter corresponds to the top dashed line that is closest to $C(t)$).

We expect the agreement between the exponent of the powerlaw decay and the generalized dimensions to improve and the powerlaw behavior to persist for longer times for system sizes larger than $L=16$. Good indication for this claim is found in Fig.~\ref{fig:D2} (a) and (b), where we compare the survival probability for $L=12$ and $L=16$ for $h=1.5$. Figure~\ref{fig:D2} (a)  shows $\langle F(t)\rangle$ and Fig.~\ref{fig:D2} (b) gives $C(t)$. It is evident that as the system size increases, the duration of the oscillations of $\langle F(t)\rangle$ and of the algebraic decay of $C(t)$ stretch out. One can see  that for $L=16$, the damping of the oscillations coincides better with $t^{-\tilde{D}_2 }$ than for $L=12$. Similarly $t^{-\tilde{D}_2^{\mbox{\tiny typ}}}$ matches $C(t)$ when $L=16$, but this is not the case for $L=12$.
\begin{figure}
\center
\includegraphics*[width=0.5\textwidth]{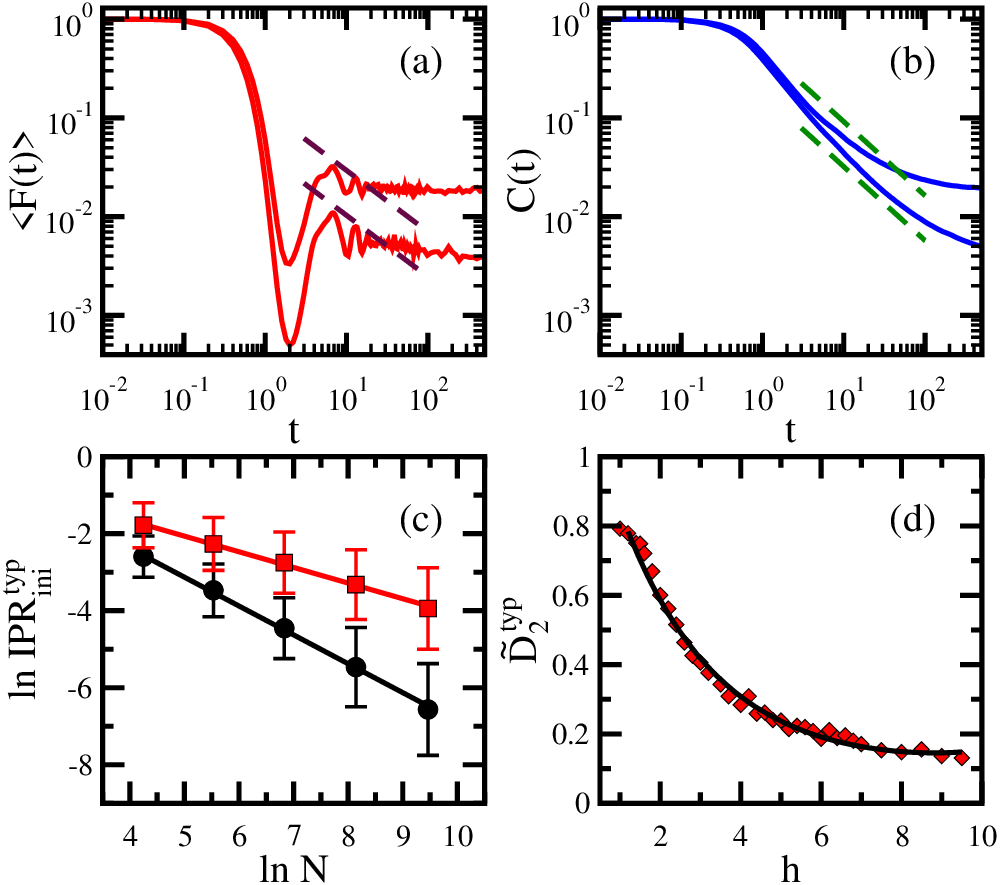}
\caption{$\langle F(t)\rangle$ (a) and $C(t)$ (b) for $h=1.5$; $L=12$ (top solid curve) and $L=16$ (bottom solid curve);  initial state is the N\'eel state. Dashed lines give $t^{-\tilde{D}_2}$ (a) and $t^{-\tilde{D}_2^{\mbox{\tiny typ}}}$ (b). For (c): Scaling analysis of $ \ln \mbox{IPR}_{\mbox{\tiny ini} }^{\mbox{\tiny typ} }$ vs $\ln {\cal N}$ for $h=1.5$ (circles) and $h=3$ (squares). Error bars are standard deviations over $10^3$ values  of $\ln \mbox{IPR}_{\mbox{\tiny ini}}$. For (d): $\tilde{D}_2^{\mbox{\tiny typ}} $ vs disorder strength. Solid line is a fitting curve. All panels: periodic boundary conditions; $\Delta=0.48$, $\lambda=0$. }
\label{fig:D2}     
\end{figure}

In Fig.~\ref{fig:D2} (c), we show two examples of the scaling analysis that we perform for $\mbox{IPR}_{\mbox{\small ini} }^{\mbox{\tiny typ} }$ to obtain $\tilde{D}_2^{\mbox{\tiny typ}}$. The slope naturally decreases as the disorder strength increases, while the standard deviations increase with it. The value of the generalized dimension becomes less precise as the system approaches the MBL phase.  

The dependence of $\tilde{D}_2^{\mbox{\tiny typ}} $ on the disorder strength is depicted in Fig.~\ref{fig:D2} (d). The decay clearly slows down for $h > 5$, but since the system sizes considered are small and finite size effects become more relevant for larger disorder strengths, we avoid speculating on this behavior.


\section{Concluding Remarks}
\label{sec:CONCLUSION}

We studied the survival probability of one-dimensional systems of interacting spins-1/2  in the absence and presence of disorder. The initial state considered was the N\'eel state. We showed that its short-time dynamics is characterized by the shape and width of the LDOS, and it does not depend on the strength of the disorder. The long-time dynamics depends on how well filled the envelope of the LDOS is. As the disorder strength increases, the LDOS becomes more sparse and the decay at long times slows down significantly. The decay becomes powerlaw with an exponent that coincides with the generalized dimension. 

We have numerical results for various observables of experimental interest, such as magnetization and spin-spin correlations. In the presence of disorder, they also show algebraic decays.
A natural extension to the present work, which we are studying, is the search for analytical expressions also for these observables. One expects the expressions obtained for the survival probability to help in this direction.

\begin{acknowledgements}
This work was motivated by a presentation given by one of the authors at the Workshop: Quantum Information and Thermodynamics held in S\~ao Carlos in February, 2015. This work was supported by the  NSF grant No.~DMR-1147430. E.J.T.H. acknowledges support from CONACyT, Mexico. LFS thanks the ITAMP hospitality, where part of this work was done.
\end{acknowledgements}



\end{document}